\documentclass[preprint]{ptephy_v1}

\preprintnumber{XXXX-XXXX} 

\usepackage{graphics}
\usepackage{natbib}
\usepackage{siunitx}
\usepackage{ulem}

\usepackage[left]{lineno}

\begin{document}

\title{Measurement of ambient neutrons in an underground laboratory at Kamioka Observatory}


\author[1,*]{Keita Mizukoshi}
\affil{Department of Physics, Graduate School of Science, Osaka University, Toyonaka, Osaka 567-0043, Japan \email{mzks@km.phys.sci.osaka-u.ac.jp }}
\author[2]{Ryosuke Taishaku}
\affil{Department of Physics, Kobe University, Kobe, Hyogo 657-8501, Japan}
\author[3]{Keishi Hosokawa}
\affil{Research Center for Neutrino Science, Tohoku University, Sendai, Miyagi 980-8578, Japan}
\author[4,5]{Kazuyoshi Kobayashi}
\affil{Kamioka Observatory, Institute for Cosmic Ray Research, the University of Tokyo, Higashi-Mozumi, Kamioka, Hida, Gifu, 506-1205, Japan}
\author[2]{Kentaro Miuchi}
\affil{Kavli Institute for the Physics and Mathematics of the Universe (WPI), the University of Tokyo, Kashiwa, Chiba, 277-8582, Japan}
\author[6,7]{Tatsuhiro Naka} 
\affil{Nagoya University, Nagoya, Aichi 464-8602, Japan}
\author[4,5]{Atsushi Takeda}
\affil{Kobayshi-Maskawa Institute for the Origin of Particles and the Universe, Nagoya, Aichi 464-8602, Japan}
\author[8]{Masashi Tanaka}
\affil{Global Center for Science and Engineering (GCSE), Faculty of Science and Engineering, Waseda University, Shinjuku, Tokyo 169-8555, Japan}
\author[9]{Yoshiki Wada} \affil{Department of Physics, Faculty of Science, Tohoku University, Sendai, Miyagi 980-8578, Japan}
\author[10]{Kohei Yorita}
\affil{Department of Physics, Waseda University, Shinjuku, Tokyo 169-8555, Japan}
\author[1,11]{Sei Yoshida}
\affil{Project Research Center for Fundamental Sciences (PRC), Osaka University, Toyonaka, Osaka 560-0043, Japan}


\begin{abstract}
    Ambient neutrons are one of the most serious backgrounds for underground experiments searching for rare events.
    The ambient neutron flux in an underground laboratory of Kamioka Observatory was measured using a $\mathrm{^3He}$  proportional counter with various moderator setups.
    Since the detector response largely depends on the spectral shape, the energy spectra of the neutrons transported from the rock to the laboratory are estimated by Monte-Carlo simulations. 
    The ratio of the thermal neutron flux to the total neutron flux was found to depend on the thermalizing efficiency of the rock.
    Therefore, the ratio of the count rate without a moderator to that with a moderator was used to determine this parameter.
    Consequently, the most-likely neutron spectrum predicted by the simulations for the parameters determined by the experimental results was obtained. 
    The result suggests an interesting spectral shape, which has not been indicated in previous studies.
    The total ambient neutron flux is $(23.52 \pm 0.68 \ \mathrm{_{stat.}} ^{+1.87}_{-2.13} \ \mathrm{_{sys.}}) \times 10^{-6}$ cm$^{-2}$ s$^{-1}$.
    This result, especially the energy spectrum information, could be a new and important input for estimating the background in current and future experiments in the underground laboratory at Kamioka Observatory.

\end{abstract}


\maketitle

\section{Introduction}\label{intro}

Ambient neutrons are one of the most serious backgrounds for underground experiments, such as neutrinoless double beta decay searches, neutrino measurements, and direct dark matter searches.
In neutrinoless double beta decay experiments, thermal neutrons can produce $\gamma$~rays close to the Q-value energy in the rock and detector components.
The $\gamma$~ray yield strongly depends on the thermal neutron flux.
In direct dark matter searches, fast neutrons can recoil target nuclei like dark matter particles.
To estimate and possibly subtract the neutron background in these experiments, a precise ambient neutron flux and spectrum are required.

Many measurements of the ambient neutron flux have been carried out in underground laboratories \cite{PhysRevD.73.053004}. 
Since the neutron energy is not directly measured by $\mathrm{^3He}$ proportional counters, which are widely used because of their large cross section to thermal neutrons, measurements with different moderator setups have been used to estimate the neutron flux in the energy ranges of interest. 
In previous studies, a simple energy spectrum consisting of a Boltzmann distribution in the thermal energy range and a $1/E$ spectrum in the high-energy range has been assumed in converting the measured count rates into a flux.
The spectral shape affects this conversion.
Therefore, an estimation of a reasonable spectrum is important.
The ambient neutron flux in the Kamioka Observatory was measured in 2002 by Minamino \cite{minamino}.
A detailed energy spectrum was not considered in that measurement.

In this paper, we considered the natural sources of ambient neutrons in the wall rock using Monte-Carlo simulations (MC) to estimate the shape of the neutron energy spectrum.
The natural sources considered were the ($\alpha$, n) reactions of $\mathrm{^{238}U}$ and $\mathrm{^{232}Th}$ chains, spontaneous $\mathrm{^{238}U}$ fission, and cosmic muons impacting the rock.
Then, the neutrons generated were transported to the laboratory.
Consequently, the most likely energy spectrum was obtained.

\section{Detector}\label{detector}

\subsection{Detection principle} 

$\mathrm{^3He}$ gas was used to detect neutrons through the following exothermal reaction,
\begin{equation}
    \label{eq:reaction}
    \mathrm{^3 He} + \mathrm{n} \to \mathrm{^3 H} + \mathrm{p} + \mathrm{0.764 \ MeV}.
\end{equation}
Information about the original kinetic energiy of an incident neutron is lost because the Q-value of the reaction, 0.764 MeV, is much larger than that of the detected neutrons.
$\mathrm{^3He}$ has a large cross section to thermal neutrons (e.g., 5333 barns at 0.025 eV \cite{crosssection}).

\subsection{Detector setup} 
The measurements were made in Lab-B at the NEWAGE \cite{newage} experimental site, one of the underground laboratories at the Kamioka Observatory.
A proportional counter (Model P4-1618-203 made by Reuter-Stokes Co.) with $\mathrm{^{3}He}$ gas at 10 atm was used.
The counter was made of a stainless steel cylinder (class SUS304), 38 cm in length and 5.18 cm in diameter.
The voltage supplied to the counter was +1300 V.

To measure high-energy neutrons (in the MeV range), moderators and a shielding material were used.
A polyethylene moderator (outer radius of 9.9 cm, length of 51 cm, and thickness of 6.5 cm) was used to thermalize the high-energy neutrons so that they can be detected by the $\mathrm{^3 He}$ proportional counter.
An additional shielding material, a 4-mm-thick boron-loaded sheet \cite{boron} (B sheet), covered the moderator to reduce the effects of ambient thermal neutrons.
The B sheet, of density 1.42 $\mathrm{g/cm^3}$, was included 20\% $\mathrm{B_4 C}$, which shields about 99.8\% of the thermal neutrons.

\begin{figure}[!t]
    \centering
    \includegraphics[width=10cm]{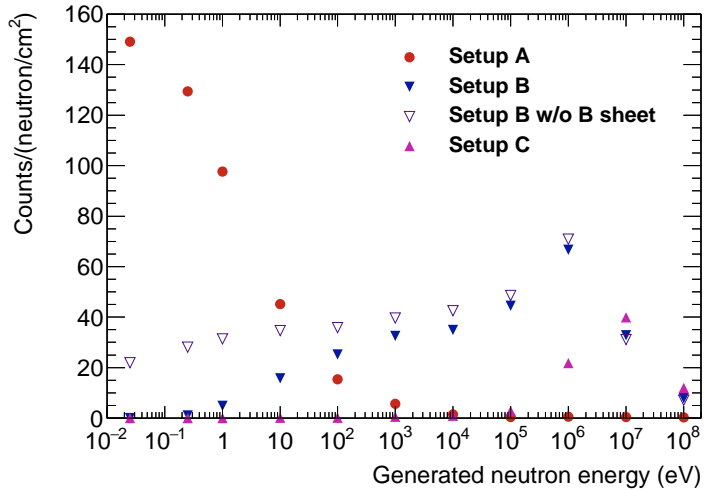}
    \caption{Estimated counts for each setup for various neutron energies with a fluence of 1~neutron/cm$^2$. The results with setup B without the boron-containing sheet (B sheet) are also shown for reference.}
    \label{fig:eff}
\end{figure}

For setup~A, there was no moderator and no B sheet.
Setup B had a moderator and the B sheet.
Setup~C was like setup B but with an additional 5-cm-thick polyethylene moderator.
The detection efficiencies in each setup were evaluated by Geant4 \cite{geant4_1,geant4_2,geant4_3}, version Geant4.10.03.patch03, with physics list QGSP\_BERT\_HP.
The geometries of the three setups were created; then, monoenergetic neutrons were generated isotropically to evaluate their responses.
Fig. \ref{fig:eff} shows the expected numbers of detected neutrons when we generated neutrons with a fluence of 1 neutrons/$\mathrm{cm^2}$.
Setups A and B are mainly sensitive to thermal and fast (--MeV) neutrons, respectively.
The simulated result for setup~B without a B sheet is also shown to illustrate the effect of the sheet.
It is found that the B sheet shields the ambient thermal neutrons making setup~B sensitive mainly to high-energy neutrons.
Setup~C was used for calibration because this setup is sensitive to the $\sim$MeV neutrons from the $\mathrm{^{252}Cf}$ calibration source as shown in Fig. \ref{fig:eff}. Details on the calibration is described in Sec. 2.3.

During the measurements with setup~A, the signal from the counter was shaped with an amplifier with a gain of 6.0 mV/fC, a rise time of 2 \si{\micro \second} and a decay time of 10 \si{\micro \second}).
The signal was recorded by a Hoshin V006 peak sensitive analog-to-digital converter.
During the measurements with setups~B and C, a different shaper with an amplification factor of 1.5 mV/fC, a rise time of 0.2 \si{\micro \second}, and a decay time of 1 \si{\micro \second} was used.
Then the signal was recorded by an Interface LPC-320910 waveform digitizer with a sampling rate of 40 MHz.
The data recorded was reduced to the pulse height and the integration of the whole pulse.
It was confirmed by the calibration that there is no defference in the signal count rate between the both readout systems of setups B and C.

\subsection{Calibration}\label{detector:calib} 

\begin{figure}[!t]
    \centering
    \includegraphics[width=10cm]{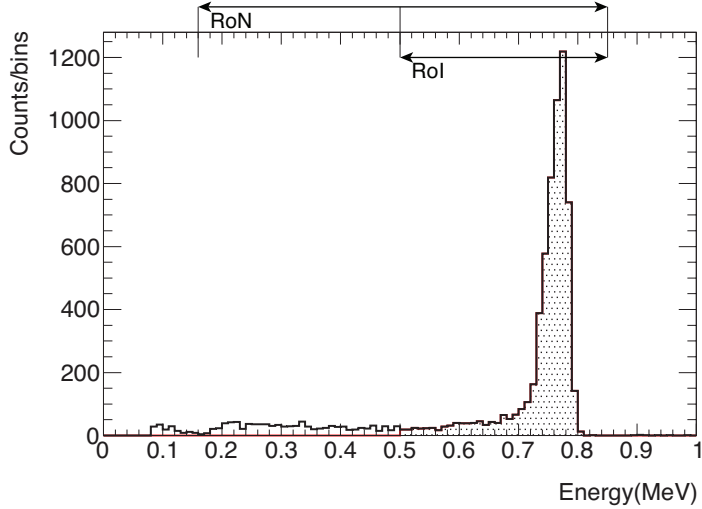}
    \caption{The energy spectrum of the $\mathrm{^3He}$ proportional counter measured during the $^{252}\mathrm{Cf}$ calibration by setup C. All the neutron events should be in the region above 0.16~MeV. The region between 0.5~MeV and 0.85~MeV was defined as the region of interest (RoI) to filter out low-energy background events.}
    \label{fig:cal}
\end{figure}

The $\mathrm{^{3}He}$ proportional counter was calibrated with a $^{252}\mathrm{Cf}$ source.
The setups B and C are sensitive to $^{252}\mathrm{Cf}$ fission neutrons with an energy of a few MeV.
Fig. \ref{fig:cal} shows the energy spectrum obtained by the $^{252} \mathrm{Cf}$ calibration.

The peak at 0.764 MeV corresponds to the Q-value in Equation (\ref{eq:reaction}).
The kinetic energies of the oppositely directed products, $\mathrm{^3 H}$ and proton, are 0.191 MeV and 0.573 MeV, respectively.
The peak occurs when the counter detects the full energies of both products.
If either product escapes out of the detector, its energy is only partly deposited.
This process is known as the wall effect for a $\mathrm{^3 He}$ counter.
The wall effect is responsible for the  flat shape below the full energy peak.

We defined a region of neutron events (RoN) between 0.16 MeV and 0.85 MeV by considering the energy resolution.
Background events, such as $\gamma$ ray, electric noise, etc., cannot be ignored in the low-energy region in the measurement of ambient neutron in the underground laboratory.
Thus, we defined a region of interest (RoI) between 0.50 MeV and 0.85 MeV to filter out these background events.
To convert the number of events in the RoI ($N_\mathrm{RoI}$) into the number in the RoN ($N_\mathrm{RoN}$), the conversion factor $\varepsilon$ was defined by the calibration data as
\begin{equation} \label{eq:ep}
    \varepsilon = \frac{N_{\mathrm{RoI}}}{N_{\mathrm{RoN}}}.
\end{equation}
From the calibration result shown in Fig. \ref{fig:cal}, $\varepsilon =  0.867 \pm 0.015$.
This conversion factor is used throughout the discussion.
Thus, the experimental count rate~$R$ was obtained using the live time $t$ and the number of events observed in the RoI, $N_\mathrm{RoI,DET}$, from

\begin{equation} \label{eq:exprate}
    R = \frac{N_\mathrm{RoI,DET}}{\varepsilon} \times \frac{1}{t}.
\end{equation}

\begin{table}[!t]
    \caption{Event rates for the measured calibration data and the estimated simulation results. The first errors are statistical. The second errors are systematic, for which only the source inner structure uncertainty was taken into account.}
\label{tab:checkMC}
\centering
\begin{tabular}{lcc}
    Setup & Calibration (cps) & Simulation (cps)\\
\hline
    Setup B & 3.31 $\pm$ 0.07 & 3.48 $\pm$ 0.05 $\pm$ 0.41 \\
    Setup C & 1.28 $\pm$ 0.02 & 1.22 $\pm$ 0.05 $\pm$ 0.15 \\
    \hline
\end{tabular}
\end{table}

This $^{252} \mathrm{Cf}$ calibration was also used to confirm the validity of the detector response simulation.
In the simulation, we created the same geometries as the experimental setups and neutrons were emitted isotropically from a $\mathrm{^{252}Cf}$ point source.
Table \ref{tab:checkMC} shows the experimental event rates for the calibration and those estimated in the simulation. 
These results are consistent within the errors.
The detector simulation, especially the thermalization in the polyethylene moderator, was confirmed by this calibration.

\section{Simulations}\label{simulation}

The energy spectrum of ambient neutrons is required to convert experimental count rates into a neutron flux.
It is extremely difficult to derive the neutron energy spectrum with a $\mathrm{^3He}$ proportional counter alone since the $\mathrm{^3He}$ counter cannot measure the incident neutron energy.
To solve this problem, we estimated the neutron energy spectrum with an MC simulation and, then, unknown parameters, {\it i.e.,} the absolute flux and thermalization efficiency, by measurements.
Neutrons are generated in the rock around the laboratory and are transported from the rock to the laboratory space.
In Sec. \ref{sim:rock}, the properties of the wall rock, such as chemical composition and radioactivity, are discussed as being common to all the simulations.
Secs. \ref{sim:an} and \ref{sim:muon} describe the neutrons generated by U/Th radioactivity and cosmic muons in the rocks, respectively.
In Sec. \ref{trans}, the energy spectra of the neutrons transported to the laboratory are shown.
Sec. \ref{shape} shows the spectral shape considering thermalization.
Finally, Sec. \ref{yield} describes the method used to derive the total neutron flux from the measured count rate and the simulated spectrum.

\subsection{Radioactivity and chemical composition of the rocks} \label{sim:rock}
Several pieces of rock were sampled from the experimental site.
The radioactivity was measured by a Ge detector.
The concentration was measured to be 0.6 ppm for $^{238}$U and 1.3 ppm for $^{232}$Th, assuming the radiative equilibrium of the U and Th series.

X-ray Fluorescence (XRF) was used to measure the chemical composition.
The rock is calc-silicate gneiss and it comprises three different types of rocks with different chemical compositions, referred to as samples 1, 2, and 3.
Sample 1 was the largest and, thus, was also assumed to be the main components of the wall rock.
Sample 2 and 3 were used to identify the effects due to the difference in chemical composition.
Table \ref{tab:chem} summarizes the results.
In addition, Table \ref{tab:chem} lists the chemical composition of two igneous rock samples that are widely distributed around Kamioka district, JR-1 and JA-3 in the geochemical reference database \cite{rock}.

\begin{table}[!t]
    \caption{Weight percentage of rock samples (insensitive to hydrogen and carbon)}
\label{tab:chem}
\centering
\small
    \begin{tabular}{lcccccccccccc}
    Sample  &$\mathrm{SiO_2}$  &$\mathrm{Al_2O_3}$ &	$\mathrm{Fe_2O_3}$ &	$\mathrm{MnO}$ &	$\mathrm{MgO}$ &	$\mathrm{CaO}$ &	$\mathrm{Na_2O}$ &	$\mathrm{P_2O_5}$ &	$\mathrm{SO_3}$ &	$\mathrm{ZnO}$ & Others\\
\hline
    Sample 1&	35.60&	11.30&	10.90&	1.08&	0.99&	39.20&	0.02&	0.35&	0.10	&0.03&	 0.43  \\
    Sample 2&	33.40&	0.73&	23.70&	4.58&	1.90&	34.00&	0.32&	0.02&	0.17&	0.17&	 1.01	\\
    Sample 3&	25.60&	0.25&	19.30&	3.73&	1.16&	41.50&	0.00&	0.02&	3.01&	5.34&	 0.09  \\
    JR-1 \cite{rock} &75.45&		12.83&	0.89&	0.10&	0.12&	0.67&	4.02&	0.02&	0.00&	0.00&	5.90\\
    JA-3 \cite{rock} &62.27&		15.56&	6.60&	0.10&	3.72&	6.24&	3.19&	0.12&	0.00	&0.00&	2.20\\
        \hline
\end{tabular}
\end{table}

\subsection{Neutrons from the uranium and thorium chains} \label{sim:an}
The main sources of neutrons in the underground laboratory are ($\alpha$, n) reactions and spontaneous fission, due to the  U/Th chains in the rock.
The yield and energy of neutrons produced by the ($\alpha$, n) reaction were calculated by NeuCBOT \cite{neucbot}.
The neutron yield was approximately 30\% different from that calculated by a widely-used similar tool, SOURCES-4C \cite{sources}.
This difference was regarded as the ambiguity of the ($\alpha$, n) simulation tools.
In addition to the ($\alpha$, n) reaction, $\mathrm{^{238} U}$ also produces neutrons through spontaneous fission.
The Watt spectrum was calculated \cite{fission}.

\begin{figure}[!t]
    \centering
    \includegraphics[width=10cm]{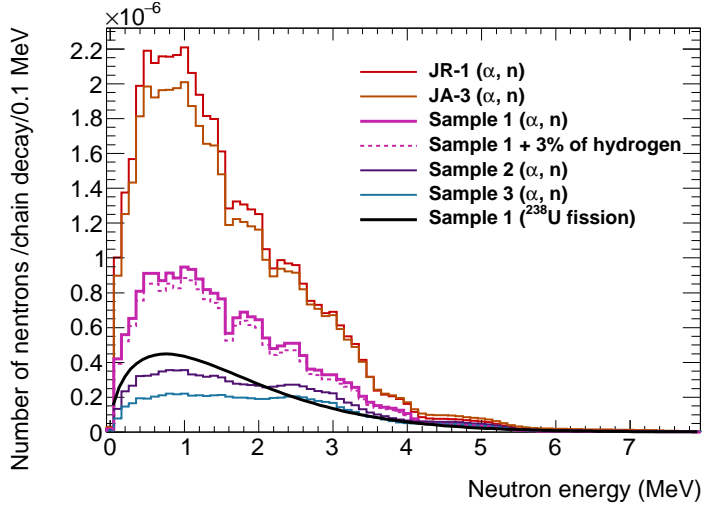}
    \caption{Simulated neutron energy spectrum generated by the $\mathrm{^{238}U}$ chain in each sample. Colored lines show the spectrum of neutrons generated by the ($\alpha$, n) reaction in each sample. The smooth black line is the spectrum due to spontaneous $\mathrm{^{238}U}$ fission (Watt spectrum). The dotted spectrum is the one with an addition of hydrogen (discussed in Sec. \ref{shape}).}
    \label{fig:an}
\end{figure}

Fig. \ref{fig:an} shows the energy spectra for the ($\alpha$, n) reactions of the U chain and for spontaneous $\mathrm{^{238}U}$ fission.
The energy spectra of the Th chain are almost same as the ones of U chain because the incident $\alpha$ energy for the ($\alpha$, n) interactions is similar to the one of U chain with an exception that no spontaneous fission is expected in the Th chain.
The number of neutrons produced changes by a factor of 10 depending on the chemical composition of the rock.
The amount of sodium, aluminum, and silicon affects the total yield.
The spectral shapes also vary with the chemical composition.
With manganese and iron, the energy spectra are likely to have high-energy components.  

\subsection{Neutrons from cosmic muons} \label{sim:muon}
\begin{figure}[!t]
    \centering
    \includegraphics[width=10cm]{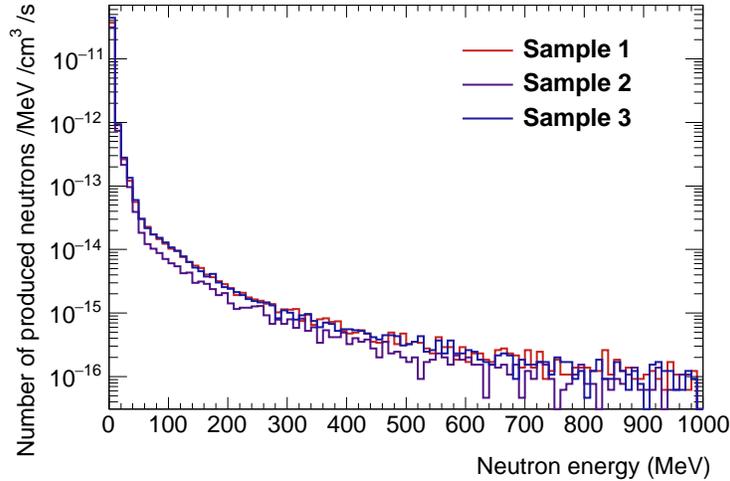}
    \caption{Simulated neutron energy spectra produced by cosmic muons for each sample from the Kamioka Observatory.}
    \label{fig:muon}
\end{figure}

Cosmic muons also generate neutrons.
The neutron energy spectrum and yield produced in the rock by cosmic muons were simulated by Geant4.
A 1-$\mathrm{m^3}$ rock cube was modeled in the simulation and muons were generated at the upper side.
The muon energy spectrum and flux at Kamioka Observatory followed the ones described in Ref. \cite{skmuon}.
In the 1-m muon path, 20--30\% of the muons produced neutrons on average.
The expected spectra of sample 1--3 are shown in Fig.~\ref{fig:muon}.

\subsection{Transportation from rock} \label{trans}
\begin{figure}[!t]
    \centering
    \includegraphics[width=10cm]{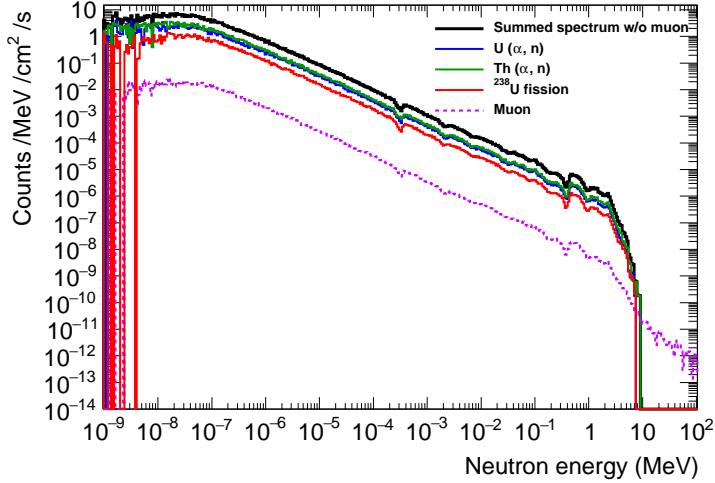}
    \caption{Transported neutron spectrum of each source for sample 1. The top black line shows summed spectrum without muon contribution. In the lowest-energy region, the statistics is low because of the limited CPU power for the simulation of thermalization.}
    \label{fig:combine}
\end{figure}

The transported neutron energy spectra in the rock were estimated by Geant4.
A 1-m-diameter sphere was placed as an experimental laboratory, surrounded by a 2-m-thick rock.
Neutrons were produced following the energy spectra shown in Fig. \ref{fig:an} isotropically from 1-m depth the rock.
Fig. \ref{fig:combine} shows the energy spectra of the transported neutron to the experimental laboratory for sample 1.
Dips around sub-keV--MeV in the spectra are due to strong resonance absorption of nuclei in the rock.

The cosmic muon can produce higher energy neutron, over 10 MeV.
The neutron yield generated from cosmic muons was about 100 times less than those from U/Th chains in the rock.
Therefore, we ignored the contribution of cosmic muons to ambient neutron flux and spectrum in this study.

\subsection{Thermalization in rock}\label{shape}  
\begin{figure}[!t]
    \centering
    \includegraphics[width=10cm]{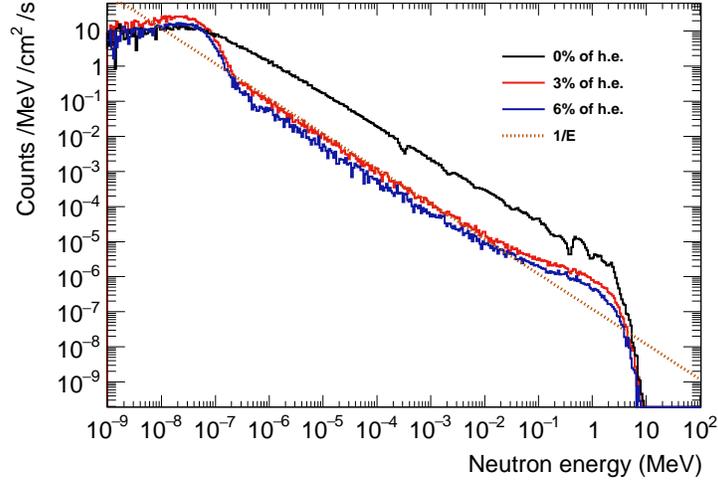}
    \caption{Transported neutron spectra from the rocks with different percentage of hydrogen equivalent (h.~e.) in sample 1. A dotted line shows $1/E$ as a flat spectrum. Thermalized spectra have excesses around 0.025 eV and 100 keV to a few MeV.}
    \label{fig:hydro}
\end{figure}

The summed spectrum without muon in Fig.~\ref{fig:combine} is not yet realistic since moderators such as hydrogen in the rock, have not been taken into account.
Thus, the deceleration and thermalization of the neutrons has not been sufficiently considered.
Any hydrogen in the rock does not significantly change the generated spectra (Fig. \ref{fig:an}) but does deform the transported spectra.
Fig. \ref{fig:hydro} shows the energy spectra from the rocks containing 0, 3 and 6\% of hydrogen by mass. 
The percentage of hydrogen was regarded as a thermalization parameter.
This parameter is referred to as \% of hydrogen equivalent (h. e.) hereafter to parametrize the thermalization in the rock.
The thermalization effect, or \% of h. e., can be experimentally determined using the ratio of the count rates of setups A and B ($R_\mathrm{A}/R_\mathrm{B}$).
As Fig. \ref{fig:eff}, since setup A is more sensitive to thermal neutrons than setup B is, a larger $R_\mathrm{A}/R_\mathrm{B}$ indicates a larger \% of h. e.
For example, $R_\mathrm{A}/R_\mathrm{B}$ values in the case of sample 1 with 0, 3, and 6\% of h. e. are simulated, and obtained to be 1.15, 2.91, and 3.29, respectively.
Therefore, it can be determined by comparing the measured and predicted $R_\mathrm{A}/R_\mathrm{B}$.
With the determined \% of h. e., the most likely spectrum is known.

The energy spectra were obtained by Geant4.
Since the thermalization process is known to be difficult to simulate, another simulation code, PHITS \cite{phits} (ver. 3.02), was also used to simulate the transportation and thermalization.
It was found that the neutrons are thermalized about 50\% more efficiently by PHITS than by Geant4 for the same \% of h. e.
For example, the spectrum with 3\% of h. e. produced by Geant4 was almost identical to that with 2\% of h. e. produced by PHITS.
The absolute amount of hydrogen in the rock can be determined with relatively large uncertainty; however, the most likely spectral shape can be obtained, regardless of the uncertainty, due to the simulation tool dependence.

\subsection{Calculation of total flux}\label{yield}  
The conversion factor to obtain the ambient neutron flux from the measured count rate was evaluated by a simulation considering the spectral shape.
Neutrons were generated with the energy spectrum obtained in Sec. \ref{shape} for the three setups.
Neutrons were beamed from a spherical surface whose radius $r$ (cm) was sufficiently large to include the setup.
The direction was weighted with a $\cos \theta$ distribution in the normal direction to realize the isotropic flux.
In this way, the fluence produced $\phi_\mathrm{MC}$ (cm$^{-2}$) is given by

\begin{equation}
    \label{eq:phimc}
    \phi_\mathrm{MC} = \frac{N_\mathrm{MC,GEN}}{\pi \times r^2} .
\end{equation}

$N_\mathrm{MC,GEN}$ is the number of neutrons generated in the simulation.
We defined the number of neutrons detected in the simulation as $N_\mathrm{MC,DET}$.
This depends on the spectral shape, as discussed in Sec. \ref{shape}.
Using the experimental count rate $R_\mathrm{A}$ (cps) for setup A, the neutron flux $\Phi$ (cm$^{-2}$ s$^{-1}$) is calculated as
\begin{equation}
    \label{eq:phir}
    \Phi = \frac{ \phi_\mathrm{MC}}{N_\mathrm{MC,DET}} \times R_\mathrm{A}.
\end{equation}

\section{Results and discussion}\label{result1}

\subsection{Experimental results}
Table \ref{tab:result} summarizes the measurements performed with the three setups.
We excluded data for which background event rate above the RoI was not stable.
About 1.5\% of the events in each run were rejected as noise events by a simple waveform analysis.
The errors of the rejection, the conversion factor $\varepsilon$, and detector gain fluctuations (within 5\%) were taken into account as systematic errors.

\begin{table}[!t]
\caption{Count rate and live time in each setup.}
\label{tab:result}
\centering
\begin{tabular}{lccccc}
    Setup &Start &Stop & Live time (day) & Rate ($\mathrm{10^{-3}}$cps)$\pm$$_\mathrm{stat.}$$\pm$$_\mathrm{sys.}$\\
\hline
    A  &Feb. 19 2016 &Mar. 20 2016  & 14.03 & 1.295 $\pm$ 0.034 $^{+0.039}_{-0.033}$\\
    B  &Oct. 19 2017 & Nov. 8 2017  &19.27  & 0.446 $\pm$ 0.018 $^{+0.013}_{-0.011}$\\
    C  & Sep. 21 2017&Oct. 19 2017& 23.97 &0.153 $\pm$ 0.009 $^{+0.005}_{-0.004}$ \\
    \hline
\end{tabular}
\end{table}

\begin{figure}[!t]
    \centering
    \includegraphics[width=10cm]{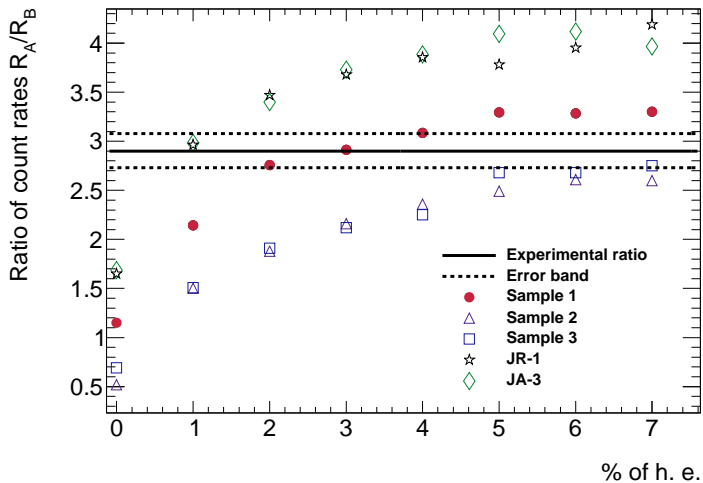}
    \caption{The ratio of count rates $R_\mathrm{A}/R_\mathrm{B}$ as a function of \% of h. e. for each rock sample. Circles, triangles, squares, stars, and diamonds show the simulated ratios for rock samples 1, 2, 3, JR-1, and JA-3, respectively. The solid line shows the experimental result, and the dotted lines show the statistical error. Sample 1 with 2--4\% of h. e. reproduces the experimental result.}
    \label{fig:hratio}
\end{figure}
\begin{figure}[!t]
    \centering
    \includegraphics[width=10cm]{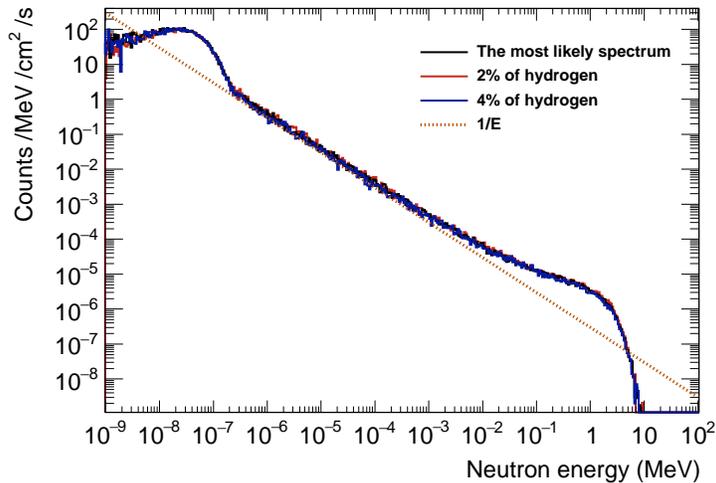}
    \caption{The most likely spectrum (sample 1 with 3\% of h. e.) of ambient neutrons produced by ($\alpha$, n) reactions and spontaneous fission in Lab-B at Kamioka Observatory.}
    \label{fig:bestfit}
\end{figure}

The ratio of the count rates for setups A and B was obtained as
\begin{equation}
    \label{eq:expratio}
    \frac{R_\mathrm{A}}{R_\mathrm{B}} = \frac{1.295 \pm 0.034 \ ^{+0.011}_{-0.010} }{ 0.446 \pm 0.018 \ ^{+0.004}_{-0.003}} = 2.90 \pm 0.14 \ ^{+0.04}_{-0.03} .
\end{equation}
Since errors due to the calibration method using $\varepsilon$ and detector inefficiency canceled out by taking the ratio, the statistical error and the systematic error from gain fluctuations were taken into account.
The measured ratio is shown by a solid line and the statistical errors by dotted lines in Fig. \ref{fig:hratio}.
The simulated ratios are also shown for different rock samples as a function of \% of h. e.
Using Equation~(\ref{eq:phir}), we obtained the total flux $\Phi$ for each energy spectrum (Table \ref{tab:rate}).
\begin{table}[!t]
\caption{Total neutron flux calculated by the simulated spectral shape, obtained from each \% of h. e. for sample 1.}
\label{tab:rate}
\centering
\begin{tabular}{lcccccccc}
    \% of h. e. & 0 &  1 & 2 & 3 & 4 & 5 & 6 & 7 \\
    \hline
    Flux $\Phi$ ($\times 10^{-6} \mathrm{cm^{-2}} \mathrm{s^{-1}}$) & 43.63& 28.50& 24.80& 23.52& 21.81& 21.81& 21.67& 21.26 
\end{tabular}
\end{table}
If we regard the amount of hydrogen is the only cause of thermalization, the most likely spectrum is obtained using sample 1 with 2 - 4\% of h. e.
The assumption, sample 1 with 3\% of h. e., well reproduces the experimental results.
The most likely spectrum assuming sample 1 with 3\% of h. e. is shown in Fig. \ref{fig:bestfit} with overlaying spectra of 2\% and 4\% of h. e. for comparison.
The differences among the three spectra are unrecognizable.
This means that this analysis is robust against the ambiguity of the \% of h. e. 

Two points (star and diamond) in Fig. \ref{fig:hratio}, JR-1 and JA-3 with 1\% of h. e., also can reproduce the experimental ratio.
These compositions derive similar spectra to the most likely one; thus, the calculated fluxes are also similar.
Their fluxes are $23.39\times 10^{-6} \mathrm{cm^{-2} s^{-1}} $ (JR-1) and $23.35\times 10^{-6} \mathrm{cm^{-2} s^{-1}}$ (JA-3) in the range of error as described below (Sec. \ref{sec:error}).

Much of the previous research assumed that the spectral shape was a Boltzmann distribution and flat ($1/E$) for thermal and fast neutrons, respectively.
The results obtained in this research basically support these assumptions but also suggest an excess in the region from 100 keV to a few MeV.
This energy range is important for dark matter experiments.
In general, our results affect the background estimation for many underground sites.

When we use the spectrum for sample 1 with 3\% of h. e., the total ambient neutron flux obtained is $23.52 \times 10^{-6} \ \mathrm{cm^{-2} s^{-1}}$ from Table \ref{tab:rate}.
Table \ref{tab:fluxes} shows the flux for each energy range.
\begin{table}[!t]
\caption{The ambient neutron flux in Kamioka Observatory.}
\label{tab:fluxes}
\centering
\begin{tabular}{lc}
    Energy range & Flux($\times 10^{-6} \mathrm{cm^{-2} s^{-1}}$) \\
    \hline
     $ <$ 0.5 eV & 7.88 \\
    0.5 eV to 1 keV & 3.11 \\
    1 keV to 1 MeV& 8.65 \\
    $\geq$ 1 MeV  & 3.88 \\
    \hline
\end{tabular}
\end{table}
Minamino reported that the flux was (8.26 $\pm$ 0.58) $\times 10^{-6} \mathrm{cm^{-2} s^{-1}}$ and (11.5 $\pm$ 1.2) $\times 10^{-6} \mathrm{cm^{-2} s^{-1}}$ for thermal ($E < 0.5$ eV) and non-thermal neutrons in Kamioka Observatory, respectively \cite{minamino}.
Minamino regarded all neutrons detected by the $\mathrm{^3He}$ counter without a moderator, like setup A, as thermal neutrons.
However, according to our study, 10--20\% of the counts for setup A in the underground laboratory were from fast neutrons. Considering this, both results on the thermal neutrons flux (7.88 $\times 10^{-6} \mathrm{cm^{-2} s^{-1}}$ in this work and 8.26 $\times 10^{-6} \mathrm{cm^{-2} s^{-1}}$ in Ref. \cite{minamino}) are consistent.

\subsection{Errors and discussion}\label{sec:error}
Table \ref{tab:error} summarizes the considered flux errors.
The error in the spectral shape is discussed in Sec. \ref{shape}.
It is estimated by the uncertainty of \% of h. e. from 2 to 4\%.
The error of detector MC for fast neutrons corresponds to the difference from $\mathrm{^{252}Cf}$ calibration as already shown in Table \ref{tab:checkMC}.
This error includes unknown inefficiency of the detector.
Since all errors are independent, we obtain $+8.5\%$ and $-9.4\%$ as the total error for the ambient neutron flux.
Consequently, the flux is $(23.52 \pm 0.68 \ \mathrm{_{stat.}} \ ^{+1.87}_{-2.13}\ \mathrm{_{sys.}}) \times 10^{-6} \ \mathrm{cm^{-2} s^{-1}}$.

\begin{table}[!t]
\caption{Errors of the ambient neutron flux}
\label{tab:error}
\centering
\begin{tabular}{lcc}
    Error & \multicolumn{2}{c}{Value (\%)} \\
    \hline
    Statistical error in measurement & \multicolumn{2}{c}{ $\pm 2.8$}\\
    Systematic error in measurement  &  +3.0 & -2.5 \\
    Spectral shape error  & +5.4 & -7.2  \\
    Error in detector MC for fast neutrons  &  +5.1 & -4.7 \\
    \hline
    Total error &  +8.5 & -9.4 

\end{tabular}
\end{table}

In principle, the simulations in this study can predict the absolute value of the neutron flux.
However, there is considerable ambiguity in the predicted flux due to uncertainties in rock properties (chemical composition, the amount of U/Th radioactivity, and density) and there is ambiguity due to the simulation tools (neutron generation and transportation). 
In this study, the total ambiguity factor was found to be more than 4.
Thus, two ambiguities in MC (total flux and \% of h. e.) were treated as unknown parameters and were determined by the experimental results.
As Fig. \ref{fig:an} shows, the spectra of neutrons generated in the rock depend on the chemical composition of the rock.
Accordingly, as Fig. \ref{fig:hratio} shows, the \% of h. e. determined by the comparison between the expected and measured $R_\mathrm{A}/R_\mathrm{B}$ values still has a large uncertainty due to the chemical composition.
However, this comparison can determine the combination of the chemical composition and \% of h. e. and the estimated spectral shapes makes very little difference.
In other words, if there are more high-energy neutrons (see samples 2 and 3 in Fig. \ref{fig:an}), then, relatively more amount of hydrogen assumption for thermalization (Fig. \ref{fig:hratio}) for a given value of $R_\mathrm{A}/R_\mathrm{B}$.
Therefore, the analysis using $R_\mathrm{A}/R_\mathrm{B}$ to obtain the spectral shape is robust against the ambiguity in the chemical composition and thermalization.
Thus, the absolute neutron flux was determined with relatively small errors.

\subsection{Sensitivity to MeV neutrons}
The spectrum and flux obtained can be used to predict the count rate for setup C.
The predicted rate, $R_\mathrm{C,MC} = (0.085 \ ^\mathrm{+0.009}_\mathrm{-0.005}) \times 10^{-3}$ cps, is smaller than the experimental one, $R_\mathrm{C} = (0.153\ \pm$ 0.009 $^{+0.005}_{-0.004}) \times 10^{-3}$ cps.
One of the reason is that $R_\mathrm{C}$ is about 10 times smaller than $R_\mathrm{A}$ and so that neutrons other than ambient ones from the rock might not be negligible.
In setup C, the contribution of fast neutrons generated around the detector, such as neutrons due to cosmic muons penetrating the detector and ($\alpha$,~n) reactions in the detector materials could have increased the count rate.
It is difficult to estimate precisely the source of fast neutrons with this detector.
Other detectors that are sensitive to higher energy neutrons, such as liquid organic scintillators, need to be used to advance our understanding of the energy spectrum predicted by this work.

\subsection{Comparison of resluts with other laboratories considering flux definition}

\newcounter{num}
\setcounter{num}{1}
    The definition of the neutron flux for this work was described in Sec. \ref{yield} (Def. \Roman{num}).
\setcounter{num}{2}
Although this definition is a standard one \cite{knoll}, there exists another definition (Def. \Roman{num}) commonly used to express the flux of particles coming from certain directions such as beams and cosmic-rays.
The flux is defined as the number of particles that passed a unit area (e.g., virtual disc, detector surface) in Def. \Roman{num}.
The result of the total neutron flux in this work is $(5.88 \pm 0.17 \ \mathrm{_{stat.}} \ ^{+0.47}_{-0.53}\ \mathrm{_{sys.}}) \times 10^{-6} \ \mathrm{cm^{-2} s^{-1}}$ with the Def. \Roman{num}.
The flux obtained by this study is almost same (20\% large) as the one measured in Modane Underground Laboratory \cite{lsm} where Def. \Roman{num} was used \cite{lsm_def}, and also consistent with the ones measured in Laboratori Nazionali del Gran Sasso \cite{gransasso} although the definition is not explicitly written.

\section{Conclusions}\label{conclusion}

Ambient neutrons are one of the most serious backgrounds for low-background experiments performed in underground laboratories.
The neutron flux and energy spectrum are required to estimate the background precisely, so that it can be subtracted effectively.
The main sources of neutrons in deep underground laboratories are ($\alpha$, n) reactions and spontaneous fission of U and Th contents in the wall rock.
An estimate of the neutron spectrum in the underground laboratory was derived by a simulation using U/Th amounts and the chemical composition of the rock as initial parameters.
The simulation requires the thermalization parameter to obtain a realistic energy spectrum.
This parameter can be determined by the experimental count rates of different setups.

The ambient neutrons were measured at Kamioka Observatory.
A $\mathrm{^3He}$ proportional counter was used to detect mainly thermal neutrons.
Higher energy neutrons were measured in different setups with different combinations of a polyethylene moderator and a B sheet.
The most likely energy spectrum was obtained.
Using the spectrum, the total neutron flux was calculated to be $(23.52 \pm 0.68 \ \mathrm{_{stat.}} \ ^{+1.87}_{-2.13}\ \mathrm{_{sys.}}) \times 10^{-6} \ \mathrm{cm^{-2} s^{-1}}$.
These experimental results and our simulations suggest there is an excess above $1/E$ in the MeV region in the ambient neutron spectrum.
The $^3 \mathrm{He}$ proportional counter is not sensitive in this region; thus, other detectors are needed to increase our detailed understanding of the spectrum structure.

\section*{Acknowledgments}

The authors would like to thank Dr. Yuji Kishimoto from KEK for lending us the $\mathrm{^3He}$ proportional counter.
We also thank Kamioka Mining and Smelting Co., Ltd. for various supports to our research activities in the underground laboratories.
We appreciate useful comments about the rock samples given by Prof. Keiko Suzuki and Dr. Koji Kiyosugi from Kobe University.
We appreciate Dr. Koichi Ichimura from the Institute for Cosmic Ray Research and Dr. Saori Umehara from Osaka University for useful advice on the rock components.
We are grateful to Dr. Yasuhiro Takemoto from Osaka University for detector operation.
This work was supported by the MEXT KAKENHI Grant-in-Aid for Scientific Research on Innovative Areas 26104001, 26104003, 26104004, 26104005, JSPS KAKENHI Grant-in-Aid for Scientific Research (S) 24224007, JSPS KAKENHI Grant-in-Aid for Scientific Research (A)16H02189, (A)17H01661.
This work was partially supported by the joint research program of the Institute for Cosmic Ray Research (ICRR), the University of Tokyo.


\end{document}